\documentclass[aps,prd,showpacs,floatfix,preprintnumbers,groupedaddress]{revtex4-1}

\usepackage{amsmath}
\usepackage{graphicx,tabularx}
\usepackage{axodraw}
\usepackage{epsfig}

\newcommand{\sla}[1]{/\!\!\!#1}

\renewcommand{\Re}{\text{Re}}

\newcommand{\nn}[1]{\tilde{\chi}^0_{{#1}}}

\newcommand{\gev}{~{\ensuremath\rm GeV}}
\newcommand{\tev}{~{\ensuremath\rm TeV}}
\newcommand{\fb}{~{\ensuremath\rm fb}}

\begin{document}

\date{\today}
\title{Observing Higgs Dark Matter at the CERN LHC}
\author{Alexandre Alves}
\email{aalves@fma.if.usp.br}
\affiliation{Instituto de F\'\i sica Te\'{o}rica, Universidade
Estadual Paulista, S\~{a}o Paulo, Brazil}

\begin{abstract} 
  Triggering the electroweak symmetry breaking may not be the only key
  role played by the Higgs boson in particle physics. In a recently
  proposed warped five-dimensional $SO(5)\otimes U(1)$ gauge-Higgs
  unification model the Higgs boson can also constitute the dark
  matter that permeates the universe. The stability of the Higgs boson
  in this model is guaranteed in all orders of perturbation theory by
  the conservation of an {\it H-parity} quantum number that forbids
  triple couplings to all SM particles. Such a unique feature of the
  model shows up as a delay in the restoration of the tree-level
  unitarity which in turn enhances the production cross section as
  compared to the standard model analogue. Recent astrophysical data
  constrain the mass of such a Higgs dark matter particle to a narrow
  window of $70$--$90\gev$ range. We show that the Large Hadron
  Collider can observe these Higgs bosons in the weak boson fusion
  channel with about 260$\fb^{-1}$ of integrated luminosity in that
  mass range. 
\end{abstract}


\maketitle


\subsection{Introduction}
\label{sec:intro}

The mechanism responsible for the breaking of the
electroweak symmetry and
the necessity of existence of cold dark matter (CDM) relics in the
universe have motivated a large theoretical effort towards a
beyond Standard Model (SM) physics in the last decades. \smallskip

The electroweak symmetry breaking (EWSB)~\cite{ewsb,djouadi} is
achieved in the Standard Model (SM) and in many of its extensions
introducing scalars whose couplings to fermions and gauge bosons
generate their masses. Within SM there remains a scalar particle in
the spectrum: the Higgs boson. 
One of the major goals of the LHC is the
detection and the study of the properties of the Higgs boson which, 
in this respect, constitutes a window for the whole EWSB
mechanism and hopefully for the high energy structure of the new
physics. On the other hand, the indirect detection of the dark matter
particle at colliders (whose existence has been established by the
WMAP experiment~\cite{wmap}) is important to establish the nature of
dark matter and, again, constitute a mean to decide between new
physics models.\smallskip

In a recent work~\cite{hosotani} an interesting connection between
these two phenomena was proposed: the Higgs boson could also be the
dark matter relic with a mass of $\sim 70\gev$ in order to satisfy the
constraints from WMAP~\cite{wmap}. In fact, the actual mass of such a
particle may lie in a narrow
$70$--$90\gev$ range as shown in Ref.~\cite{cheung2}. This connection
has a deep impact 
in the Higgs boson phenomenology at colliders because the Higgs boson
becomes absolutely stable in the warped 5D $SO(5)\otimes U(1)$
gauge-Higgs model proposed  in Ref.~\cite{hosotani}. 
As a consequence of the conservation of a new
quantum number which forbids all triple couplings to SM particles 
(only the quartic couplings to weak bosons and fermions remain) the
Higgs boson acquires a property shared by {\it R-parity} conserving
particles of the Minimal Supersymmetric Standard Model
(MSSM)~\cite{mssm} and {\it KK-parity} conserving particles from
Universal Extra Dimensions (UED)~\cite{ued} scenarios for example --
they are produced at colliders exclusively in pairs.\smallskip
 
In Ref.~\cite{cheung1} the double Higgsstrahlung
channel $pp\rightarrow (W,Z)HH$ was considered at the LHC as a
possible detection channel but the backgrounds were found to be three orders of
magnitude larger than the signal which would demand an extremely large
amount of data for detection. Associated production of Higgs and top or
bottom quark is possible though $f\bar{f}HH$ is a dimension-five operator
suppressed by a factor $m_f/v^2$ where $v=246\gev$ is the Higgs
{\it vev}. Top quark and weak boson pair production
can receive contributions 
from quartic Higgs boson couplings but at one-loop level. Moreover,
these kinds of processes must be initiated by bottom partons once
the Higgs couples to mass.\smallskip

In Ref.~\cite{eboli} the potential of the LHC to discover Higgs bosons
decaying into invisible particles
like gravitons, gravitinos or neutralinos is demonstrated in the weak
boson fusion (WBF) channel $pp\to jjH\to jj\sla{E}_T$. In this case
the far forward tagging jets work as an experimental trigger at the
same time they endow the signal events singular features that make the
separation from background events possible. The discovery potential of
the LHC in the WBF channel has been established for a number of SM
Higgs boson decaying channels~\cite{wbf:sm}, Higgs bosons of the
MSSM~\cite{wbf:mssm}, and in other models~\cite{wbf:others}.\smallskip

Being electrically neutral, weakly interacting, and absolutely stable
 the detectors would miss all Higgs signals -- only indirect detection
 would be possible 
via large missing momentum topologies, a typical dark matter signal at
colliders. Such signals need additional charged leptons or jets 
which can serve as an
experimental trigger just like the invisibly decaying Higgs case of
Ref.~\cite{eboli}. We thus propose the weak boson fusion
\begin{equation}
pp\rightarrow jjHH\rightarrow jj\sla{E}_T
\label{eq:ppjjhh}
\end{equation}
as a potential searching
channel at the LHC. The two far forward tagging jets working as the
required trigger plus the large amount of missing momentum
associated to the stable Higgses constitute our signal. The lack of
triple couplings to SM gauge bosons delay the restoration of unitarity
which effectively occurs only when the heavy ${\cal O}(1)\tev$ KK
gauge bosons associated to the larger gauge group $SO(5)\otimes U(1)$
start to propagate~\cite{haba} resulting in
an enhanced cross section compared to the SM case as we will discuss in
section~\ref{unitarity}.\smallskip 

In summary in this work we investigate the Higgs pair production of a
gauge-Higgs unification model with a high KK mass scale as
proposed in Ref.~\cite{hosotani} in the
weak boson fusion channel at the $14\; \hbox{TeV}$ LHC in the
$70$--$90\gev$ mass range consistent with the PAMELA, HESS,
and Fermi/LAT astrophysical data as shown in Ref.~\cite{cheung2}.
Moreover, contrary the SM
case~\cite{djouadi,haba}, cancellations among triple and quartic
Higgs contributions do not take place until the heavy KK states start
to propagate which results in an enhanced 
production cross section as we discuss in
section~\ref{sec:model}. As in the 
case of the single Higgs production~\cite{eboli}, the two 
far forward tagging jets serve as triggers and as an efficient tool to
reduce the SM backgrounds. 
We show that with sufficient integrated luminosity the Higgs dark
matter particle of this model can be observed at the CERN LHC.


\subsection{The Effective Lagrangian of the $SO(5)\otimes U(1)$
  Gauge-Higgs Unification Model}
\label{sec:model}

The model we are considering is a gauge theory based on the group
$SO(5)\otimes U(1)$ with an warped extra
dimension~\cite{hosotani}. The 4D Higgs boson is identified with a
part of the extra-dimensional component of the gauge bosons. As a
consequence the
Higgs couplings to all particles are fixed by the gauge principle. If
the extra-dimension space is non--simply--connected the
4D neutral Higgs boson will correspond to quantum fluctuations of an
Aharonov-Bohm phase $\theta_H$ which is a physical degree of freedom
originated from vanishing field strengths $F_{MN}=0$.\smallskip

Quantum corrections generate an effective Higgs potential
$V_{eff}(\theta_H)$ which depends on the Aharonov-Bohm phase
$\theta_H$ along the fifth dimension. It has been shown that the
effective potential has a global minimum at
$\theta_H=\pm\pi/2$~\cite{hosotani}. Small fluctuations around
the minimum $\theta_H=\pi/2+H(x)/v$ introduce the Higgs field
$H$ responsible for the electroweak symmetry breaking. 

Mirror reflection symmetry in the fifth dimension and the invariance
under transformations of the larger group implies that 
the effective Higgs
potential must be invariant under $H\rightarrow -H$. The Higgs field
is odd under the {\it H-parity} transformations while all other SM
particles remain even.\smallskip

For our studies the relevant effective four dimensional interaction Lagrangian of the
$SO(5)\otimes U(1)$ gauge theory in the 5D warped spacetime
incorporating all those symmetries is given by 
\begin{eqnarray}
{\cal L}_{eff} & = & \frac{g^2 v}{2}\cos(\theta_H)W_\mu W^\mu
H+\frac{g^2 v}{8c^2_W}\cos(\theta_H)Z_\mu Z^\mu
H\nonumber \\
&+& \frac{g^2}{4}\cos(2\theta_H)W_\mu W^\mu HH 
+\frac{g^2}{8c^2_W}\cos(2\theta_H)Z_\mu Z^\mu HH\nonumber \\
&+& \sum_f
\frac{m_f}{2v^2}\cos(2\theta_H)\bar{\psi}_f\psi_f HH 
\label{eq:leff}
\end{eqnarray}

As a consequence of the conservation of the {\it H-parity } all triple
vertices involving Higgs interactions vanish at $\theta_H=\pi/2$. 
Only quartic interactions
with SM strength survive and the Higgs boson becomes absolutely
stable.\smallskip

Note that the vanishing of $ZZH$ coupling evades the LEP bound on the
Higgs mass. As a matter of fact, all possible collider constraints on
the Higgs mass are evaded. Nonetheless, mass bounds from WMAP,
PAMELA, HESS, and Fermi/LAT allow a narrow $70-90\gev$~\cite{cheung2}
window for collider searches. \smallskip

\subsection{Search strategy at the CERN LHC and Calculational Tools}
\label{sec:search}

The two far forward hard jets plus missing energy topology 
has already been studied in the literature at the leading order for a
single invisibly decaying Higgs boson production in weak boson
fusion~\cite{eboli}. The backgrounds are exactly the same and the
whole analysis can be done along the same lines. For this reason we
use the results and discussions for the backgrounds analysis from that
work and simulate only the signal events for $pp\rightarrow
jjHH\rightarrow jj\sla{E}_T$.\smallskip

The backgrounds consist of process leading to two jets and missing
transverse momentum: (1) QCD and WBF $Zjj\rightarrow jj\nu\bar{\nu}$,
(2) QCD and WBF $Wjj\rightarrow jj\ell\nu$ where the charged lepton is
not identified, (3) QCD multijet production with large missing
momentum generated by energy mismeasurements or high transverse
momentum particles escaping detection through the beam-hole.
\smallskip

\begin{figure}[t]
  \begin{center}
  \includegraphics[width=9.5cm]{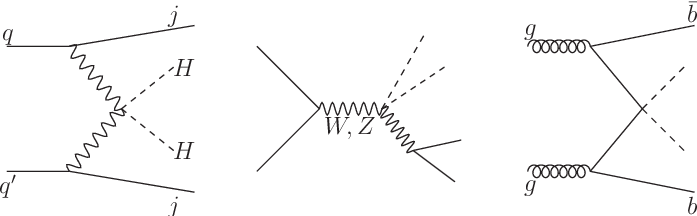} 
  \end{center}
\vspace{-2mm}
\caption[]{Feynman diagrams contributing to the $pp\to jjHH$
  process. From left to right we show the WBF, double Higgsstrahlung,
  and the gluon fusion QCD contribution respectively. There are
  additional QCD diagrams not shown in the figure initiated by
  $q\bar{q}$, $gb(\bar{b})$, and $b\bar{b}$.}
\label{fig:feynman}
\end{figure}

Our signal was simulated at parton level with full leading order tree level matrix
elements and full CKM matrix using the Calchep package~\cite{calchep}
including electroweak and QCD contributions. The electroweak
contributions consist of genuine WBF diagrams and double
Higgsstrahlung diagrams with a $Z$ or $W$ boson decaying into jets as
can be seen at Figure~(\ref{fig:feynman}). The QCD contributions are
initiated by gluons, light or bottom partons and the Higgses are
radiated off the final state bottom quarks. The background events were
simulated at parton level with full tree level matrix elements using
the Madevent package~\cite{madevent} (see Ref.~\cite{eboli}).\smallskip
NLO QCD corrections to the Higgs pair production in WBF process were calculated
at~\cite{figy1} and a very modest K-factor around 1 was found.

We do not take into account the contributions from the heavier KK-modes of
gauge bosons since 
we expect they will unitarize the scattering amplitudes only at 
some higher scale as discussed in section~\ref{unitarity}. Instead we
used the unitarization prescription suggested at Ref.~\cite{wwstrong}
in order to mimic the effect of the heavy KK gauge bosons at the
${\cal O}(1)\tev$ scale relevant for the LHC phenomenology. Yet the WBF cross
section for the model under consideration is expected to be larger than the SM analogue.\smallskip

In order to be consistent with the simulations performed in
Ref.~\cite{eboli} we
employed CTEQ4L parton distribution functions~\cite{cteq} at the
factorization scale $\mu_F =\min(p_T)$ of the defined jets. The
electroweak parameters $\sin^2\theta_W=0.23124$,
$\alpha_{em}=1/128.93$, $m_Z=91.189\; \hbox{GeV}$, and $m_W=79.95\;
\hbox{GeV}$ were taken from Ref.~\cite{eboli} as well. We simulate
experimental resolutions by smearing the energies (but not directions)
of the defined jets with a Gaussian error given by $\Delta
E/E=0.5/\sqrt{E}\oplus 0.02$ ($E$ in GeV).\smallskip 

We checked that all kinematic distributions used to impose cuts on the
final states jets and missing momentum in our case are very similar
to those of $pp\rightarrow jjH$ of Ref.~\cite{eboli}. Therefore, we
can assume the same strategy to suppress the backgrounds and impose the
same cuts 
\begin{alignat}{7}
p_{T_j} &> 40 \gev  \qquad
&|\eta_j| &< 5.0 & \notag \\
|\eta_{j_1}-\eta_{j_2}| &> 4.4 \qquad
&\eta_{j_1}\cdot\eta_{j_2} &< 0 \notag \\
\sla{p}_T &> 100 \gev 
&m_{jj} &> 1200 \gev & 
\label{eq:cuts}
\end{alignat}
%







The cuts on the transverse momentum and rapidity of the defined jets
are the standard WBF selection cuts while the missing momentum cut
explores the features of our signal with two final state dark
matter particles. A very large cut on the invariant mass of the two
tagging jets, $m_{jj}$, helps to reduce the backgrounds further. \smallskip

The jets from the double Higgsstrahlung contributions (see
Fig.~(\ref{fig:feynman})) are on the $W$ or $Z$ mass shell being
eliminated after the jets invariant mass cut is applied remaining only
the genuine WBF contributions. The QCD contributions:
$gg(q\bar{q})\rightarrow b\bar{b}HH$, $b\bar{b}\rightarrow ggHH$,
$gb(\bar{b})\rightarrow gb(\bar{b})HH$ involve double Higgs radiation
from the bottom quark lines which are suppressed by $m_b/2v^2$ (see
Eq.~(\ref{eq:leff})).  Moreover, the jets from this class of
contributions are typically central and softer than the jets from WBF
contributions being negligible after imposing the cuts. The large SM
QCD backgrounds are strongly reduced after the missing momentum cut is
applied. A survival probability associated to a central soft jet
activity veto of $0.28$ for QCD processes, as estimated
in~\cite{rainwater} and used in Ref.~\cite{eboli}, is crucial to
reduce these backgrounds and enhance the signal to background
ratio. Yet they are ineffective against the SM WBF processes whose
survival probability is high: $0.82$. \smallskip

For this reason the dominant background contribution after the cuts~(\ref{eq:cuts})
is the Standard Model WBF contribution~\cite{eboli}. To suppress this background
the following cut on the the azimuthal angle between the tagging jets, $\phi_{jj}$,
was proposed in Ref.~\cite{eboli}\smallskip
\begin{equation}
\phi_{jj} < 1
\label{eq:cutphi}
\end{equation}

It explores the nature of the particle being produced between the
tagging jets: a scalar (Higgs) for the signal and a vector ($W,Z$
boson) for the background. The spin of this particle determines the
angular distributions of the tagging jets and eventually of the products
of their decays. \smallskip

Next we discuss the tree-level unitarity violation in the $WW\to HH$
scattering and unitarization prescription used to compute a
reliable cross section for the $pp\to jjHH$ process.

\subsection{Tree-Level Unitarity Violation in $WW\to HH$ Scattering}
\label{unitarity}

\begin{figure}[t]
  \begin{center}
  \includegraphics[width=9.5cm]{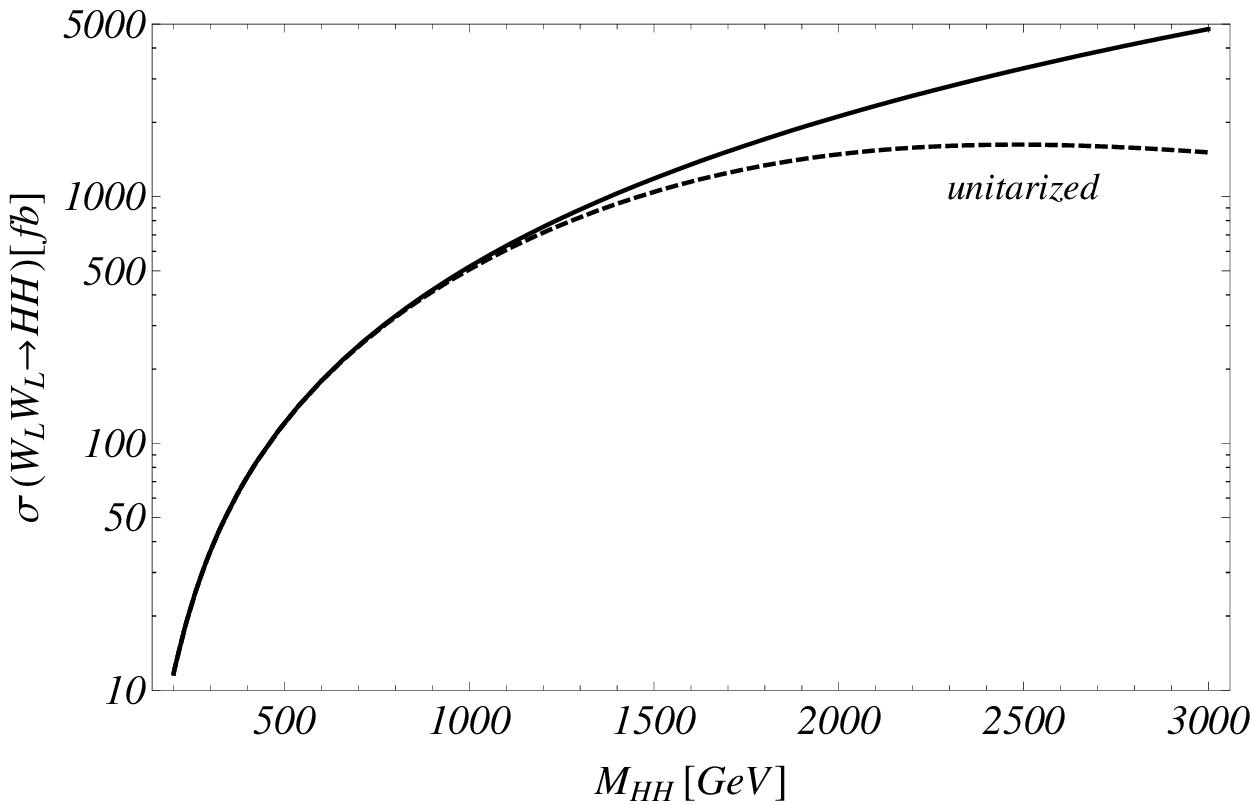} 
  \end{center}
  \vspace*{-8mm}
  \caption{The $W_LW_L\to HH$ cross section as function of the Higgs
    pair invariant mass. The solid line shows the non-unitarized cross
  section and the dashed line the cross section unitarized according
  to the elastic scattering prescription discussed in the text.}
\label{fig:xsec_uni}
\end{figure}

The lack of the triple couplings $(W,Z)HH$ and $HHH$ causes the
violation of tree-level unitarity of the $WW\to WW$ scattering at
${\cal O}(1)\tev$~\cite{haba}. 
The scattering amplitudes take the 
largest values precisely at $\theta_H=\pi/2$ in warped spacetimes in
these classes of models as shown in Ref.~\cite{haba}.  The ${\cal
  O}(E^2)$ terms which cause the unitarity violation are cancelled as
soon as the heavier KK-modes of gauge bosons start to propagate
leading to a constant behavior of the amplitudes as a function of the
energy $E$ at a scale around ${\cal O}(10)\tev$~\cite{haba} if the KK
mass scale is $m_{KK}\gtrsim 1\tev$. That is the
reason why we expect the double Higgs dark matter production has a
much larger cross section in the weak boson fusion channel as compared
to the SM case at ${\cal O}(1)\tev$ scale. 
The WBF process is intimately related to the EWSB and
the violation or the delay in the restoration of tree-level unitary of
the $WW\to WW$ scattering is a sign that new physics must come into
play~\cite{han_wbf}.\smallskip

In the case of $W_LW_L\to HH$ scattering the S-wave
amplitude gives the dominant contribution for the energies of interest 
\begin{equation}
a_0 = \frac{1}{64\pi}\int_{-1}^1 -i{\cal M}(W_LW_L\to HH)d\cos\theta =
\frac{\alpha_{em}}{32\sin^2\theta_W m^2_W} (M_{HH}-2m^2_W)
\end{equation}
where $\theta$ is the polar angle of the scattered Higgs boson in the
lab frame, and $M_{HH}=(p_H+p_H)^2=(p_{W^+}+p_{W^-})^2$ is the $HH$
center-of-mass energy square. Unitarity of the scattering amplitude
requires that $|\Re\; a_0|\leq 1/2$ which in turn implies
\begin{equation}
M_{HH}< 22m_W\approx 1.8\tev
\end{equation}
in agreement with the scale at which $W_LW_L\to W_LW_L$ unitarity is violated
as shown in~\cite{haba}.\smallskip

The growth in the scattering cross section as a function of
$M_{HH}$ for $W_LW_L\to HH$ can be seen from the solid line
of Fig.~(\ref{fig:xsec_uni}). 
As soon as the KK modes of the heavy
gauge bosons start to propagate the cross section should cease to increase
becoming flat at very high energies. 
The dashed line shows the regularized cross section as a function
of $M_{HH}$ using the
elastic scattering prescription proposed in Ref.~\cite{wwstrong}
replacing $a_0$ by $a_0/(1-ia_0)$ in the calculation of the
amplitude in order to mimic the regularizing effect of the KK states.\smallskip

\begin{figure}[t]
  \begin{center}
  \includegraphics[width=8.5cm]{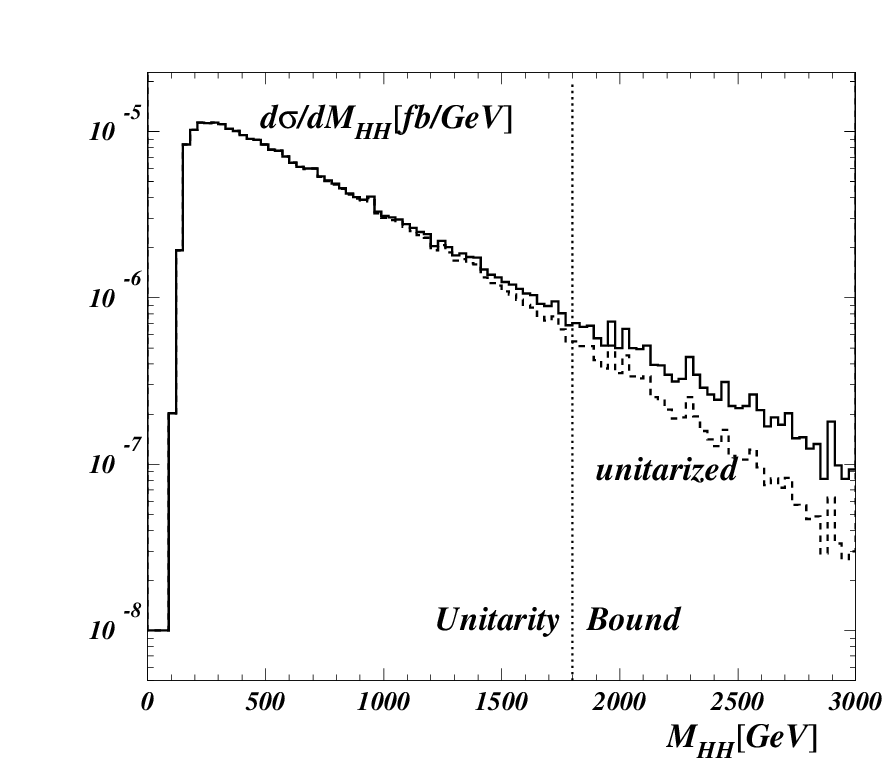} 
  \end{center}
  \vspace*{-8mm}
  \caption{The Higgs pair invariant mass distribution for the process
    $pp\to jjHH$. The solid histogram represents the non-unitarized
    distribution and the dashed histogram the unitarized one.
    Only 5\% of all events lies above the unitarity
    violation threshold scale of $1.8\tev$.}
\label{fig:mhh}
\end{figure}

However if the KK mass scale is
expected to be of order 1 TeV or higher, the impact on the $W_LW_L\to
HH$ should be small once the phase space to produce heavy on-shell KK states is
restricted. To confirm that expectation we show at the 
Fig.~(\ref{fig:mhh}) the invariant mass distribution for the Higgs
pair in the $pp\to jjHH$ process. We see that $d\sigma/dM_{HH}$ peaks
around $300\gev$ only being very small at $2m_{KK}\sim 1.8\tev$. In fact 
only 5\% of all events present a Higgs pair invariant mass above the
unitarity threshold of $1.8\tev$.
Yet their contribution to the scattering amplitude might be important near
the violation threshold energy. The dashed histogram at Fig.~(\ref{fig:mhh})
represents the unitarized distribution confirming our expectation of
the small impact of the heavy KK states at the energy scale relevant
for our studies. After unitarization the total cross section decreases
by a modest factor of 0.965 which was taken into account in the subsequent analysis.  
\section{Results and Discussions}
\label{sec:results}

The signal total cross section without cuts is $29.05(27.3)\; \hbox{fb}$ for a
$70(90)\gev$ Higgs boson at the $14\;\hbox{TeV}$ LHC while the total
background cross section amounts to 
$2.79\; \hbox{pb}$. As we anticipated in section~\ref{sec:model} the
$pp\to jjHH$ rate in this gauge-Higgs 
unification model is about 5 times larger than its SM analog whose
WBF production cross section for 
a $70\gev$ Higgs mass is $6.3\fb$. The same set of parameters,
factorization scale and parton distribution functions for the Higgs
dark matter case were used in this computation.\smallskip
\begin{table}[ht]
 \begin{tabular}{|c||c||c||c|}
\hline \hline
$\sigma$(fb) &  signal: $m_H=70(90)\gev$  & Total Background & ${\cal L}$ (fb$^{-1}$)
\\
\hline
 basic cuts (bc) &  9.7(9.3)   & 918 & 246(267)
\\
\hline
 bc + $\phi_{jj}<1$ &  4.05(4.03) &  167 & 255(257)       
\\
\hline \hline
 \end{tabular}
 \caption{Signal and background cross sections after basic
   cuts~(\ref{eq:cuts}) and basic cuts plus the $\phi_{jj}$ cut of
   Eq.~(\ref{eq:cutphi}). The survival probability after a soft
   central jet veto is incorporated already as discussed in
   section~\ref{sec:search}.  
   We show the results for a $70$ and a
   $90\gev$ Higgs boson. The last column displays the required integrated
   luminosity for a $5\sigma$ significance observation.}   
\label{tab:results}
\end{table}

After applying all the basic cuts~(\ref{eq:cuts}) and the $\phi_{jj}$
cut, and assuming the same survival probability for a central soft jet veto
of $P_{surv}=0.87$ for our signal as in Ref.~\cite{eboli}, we found
$4.05(4.03)\; \hbox{fb}$ for the signal against $167\; \hbox{fb}$ for the
total background for a $70(90)\gev$ Higgs boson mass. 
Based on these rates a $5\sigma$ significance observation is
possible with $255(257)\hbox{fb}^{-1}$ of integrated luminosity for a
$70(90)\; \hbox{GeV}$ Higgs dark matter particle. The
table~\ref{tab:results} summarizes our results.\smallskip

\begin{figure}[t]
  \begin{center}
  \includegraphics[width=8.5cm]{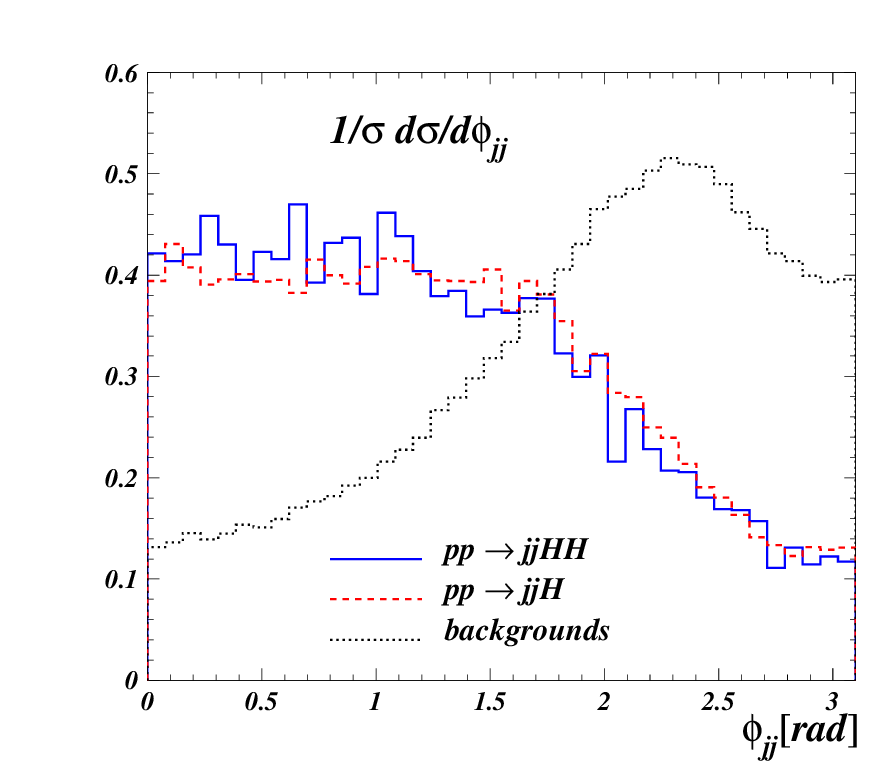}
  \end{center}
  \vspace*{-8mm}
  \caption{The normalized distributions of the azimuthal angle
    separation between the tagging jets for the double Higgs ($pp\to
    jjHH$), single Higgs ($pp\to jjH$), and total background after
    applying the cuts~(\ref{eq:cuts}).}
\label{fig:phijj}
\end{figure}

We show in Figure~(\ref{fig:phijj}) the normalized distributions for
the $\phi_{jj}$ variable for our double Higgs
signal, the single invisibly decaying Higgs, and the total background
after applying the cuts from Eq.~(\ref{eq:cuts}). The effect of the
spin nature of the particles being produced is evident from this plot
and motivates the cut~(\ref{eq:cutphi}) devised to suppress the SM
weak boson fusion background. \smallskip

The impact of the $\phi_{jj}$ cut~(\ref{eq:cutphi}) on the double Higgs
production is fairly the same as the single Higgs process representing
a dilution factor of $0.37$ in our case and $0.40$ in the single Higgs case.
The backgrounds on their turn are suppressed by a factor of $0.18$ which
motivates the cut in the single Higgs case after all. The impact of
this cut on the required luminosity is mild though changing the
required integrated luminosity for a $5\sigma$ observation from
$246(267)\fb^{-1}$ to $255(257)\fb^{-1}$ to a $70(90)\gev$
Higgs. \smallskip

As suggested in
Ref.~\cite{eboli}, an analysis based on the shape of the $\phi_{jj}$
distribution could be a good idea to separate signal from backgrounds
but the reduced signal cross section after cuts in our case is a
challenge for that purpose.\smallskip

It indicates that there is still some room for an
optimum choice of cuts for example tightening the missing $p_T$ or
the jets invariant mass cut. On the other hand a complete simulation
taking into account hadronization, pile-up effects and more realistic
detector efficiencies is necessary in order to evaluate the LHC
potential more precisely.\smallskip

An interesting question arisen looking at the $\phi_{jj}$ distribution
of Figure~(\ref{fig:phijj}) is how to discriminate between a Higgs dark
matter of the classes of models under consideration 
and a single Higgs decaying to an invisible final state with a
large branching ratio as considered in Ref.~\cite{eboli}. 
It is evident it cannot be done based on the
kinematic distributions of the jets. Observing signals associated to a
single invisible Higgs boson production in other channels would be
important for that task 
once we do not expect seeing signals for the classes of models we
considering here based on our results and on the results from
Ref.~\cite{cheung1} as well. \smallskip

Another possibility is to probe the high energy growth of the
longitudinal vector boson scattering using dedicated techniques for
hadron colliders as those proposed in Ref.~\cite{han_wbf,butter} for
example.  
A future linear collider could do that job easily in the WBF channel
once the energy of the incoming leptons can be tuned. 

It's worth mentioning that dark matter production in weak boson fusion
is not likely to occur at hadron colliders in the framework of models
presenting CDM candidates like the
MSSM or UED for example. In the MSSM case it has been shown~\cite{plehn} that
destructive interference between WBF-type diagrams and bremsstrahlung
diagrams contributing to
the lightest neutralino production decreases the rates to an 
attobarn level for $pp\to jj\nn{}\nn{}$ at the LHC. 
On the other hand in the UED models the
lightest KK particle is an almost pure $U(1)_Y$ gauge boson which
strongly suppresses the couplings to other gauge bosons and probably
depletes the WBF channel as well. \smallskip
\section{Conclusions}
\label{sec:conclusions}

In the $SO(5)\otimes U(1)$ gauge-Higgs unification model in the
Randall-Sundrum spacetime proposed in Ref.~\cite{hosotani}
 the 4D Higgs field becomes a part of the
fifth-dimensional component of the gauge fields. Electroweak
symmetry is broken dynamically through loop corrections to the Higgs
potential and the conservation of an additional quantum number called
{\it H-parity} renders absolute stability to the Higgs boson which
becomes a natural cold dark matter candidate.  As a consequence of
vanishing of all triple couplings to the Standard Model spectrum by
virtue of {\it H-parity} conservation, the Higgs boson can only be
produced in pairs through quartic couplings to massive gauge bosons
and fermions. 
The mass of such a Higgs dark matter is constrained from recent
astrophysical data to lie in the $70$-$90\gev$ mass range. \smallskip

In this paper we show that for an ${\cal O}(1)\tev$ KK mass scale
the $14\; \hbox{TeV}$ LHC has the potential 
to discover such a Higgs dark matter particle of $70(90)\gev$ mass
with $255(257)\fb^{-1}$ of integrated luminosity for a $5\sigma$
observation 
in the weak boson fusion channel following the same search strategy
used in the single invisibly decaying Higgs case of
Ref.~\cite{eboli}. On the other hand, the cut on the azimuthal angle
between the tagging jets proposed in Ref.~\cite{eboli} to reduce the
SM WBF backgrounds were found to be less effective from the 
point of view of reducing the amount of data necessary for discovery
compared to the single Higgs case which demonstrates that there is
still room for optimization of the search strategy. \smallskip

As a final comment, the enhanced production cross section compared to
the SM case is due the lack of cancellations between the triple and
quartic contributions to the double Higgs production. This feature may
help to establish the model if the 
growth of the vector boson scattering amplitudes as a function of the
energy could be determined for example in a future
linear collider or in the LHC using dedicated methods to that aim as for 
example was proposed in~\cite{han_wbf}.


\begin{acknowledgments}
  We would like to thank Oscar \'Eboli for reading the manuscript and
  encouragement. This research was supported 
  by Conselho Nacional de Desenvolvimento Cient\'{\i}fico
  e Tecnol\'ogico (CNPq).
\end{acknowledgments}




\end{document}